\documentclass[fleqn,10pt]{wlscirep}
\usepackage[utf8]{inputenc}
\usepackage[T1]{fontenc}
\usepackage{subcaption}
\usepackage{multirow}

\title{Reinforcing Prestige: Journal Citation Biases in Astronomy}

\author[1,2,*]{Vardan Adibekyan}
\author[1,2]{Olivier Demangeon}
\author[1,2]{Tiago Campante}
\author[1,2]{Nuno Santos}
\author[1,2]{Susana Barros}
\author[3]{Artur Hakobyan}
\affil[1]{Instituto de Astrof\'isica e Ci\^encias do Espa\c{c}o, Universidade do Porto, CAUP, Rua das Estrelas, 4150-762 Porto, Portugal}
\affil[2]{Departamento de F\'{\i}sica e Astronomia, Faculdade de Ci\^encias, Universidade do Porto, Rua do Campo  Alegre, 4169-007 Porto, Portugal}
\affil[3]{Center for Cosmology and Astrophysics, Alikhanian National Science Laboratory, 2 Alikhanian Brothers Str., 0036 Yerevan, Armenia}

\affil[*]{vadibekyan@astro.up.pt}

\begin{abstract}
Citations are essential for recognizing scientific contributions, yet citation behavior is shaped by more than just relevance or quality. We analyzed approximately 255,000 refereed astronomy articles published between 2000 and 2025 to investigate how journals are cited relative to their publication volume and authorship context. We find that multidisciplinary journals receive disproportionately more citations, up to nine times higher than their share of articles, while field-specific journals are cited less frequently in proportion to their output. Citations to a journal also increase significantly when authors publish within it, a bias particularly pronounced in multidisciplinary journals. Although this effect has declined over the past decade, it remains notable. These patterns likely arise from a combination of topical clustering, institutional/individual publishing habits, and strategic referencing to align with editorial expectations. Our findings reveal persistent structural biases in scientific visibility and suggest that citation-based metrics should be used with greater awareness of the publishing context they reflect. We encourage authors, reviewers, and editors to remain mindful of these dynamics and strive for fairness and inclusivity when selecting references.
\end{abstract}

\begin{document}

\flushbottom
\maketitle
%
%
\thispagestyle{empty}

\noindent\textit{This manuscript presents the full version of a shorter Research Note submitted to RNAAS.}
\vspace{0.3cm}

Citations are the primary means by which scientific work acknowledges and builds upon previous research. Ideally, references should be selected with fairness and objectivity, giving proper credit to all relevant contributions. In practice, however, unconscious biases often shape citation choices. Citation bias is not merely an analytical concern but an ethical one, with implications for the equitable dissemination of knowledge and the marginalization of underrepresented voices in science \cite{Mattiazzi-24}. While bibliometric indicators can support research evaluation, their misuse as simplistic proxies for scientific excellence or career merit can be misleading and even harmful \cite{Neylon-22}.

Many researchers have experienced the frustration of seeing their own work or that of colleagues overlooked in articles addressing topics directly related to their expertise. The authors of this manuscript are no exception. Whether such omissions stem from a lack of awareness or from conscious decisions, the result can be perceived as ignorance of the field or even deliberate exclusion. But how widespread are these citation biases? Are they isolated incidents involving individual researchers or research groups, or do they reflect systematic tendencies within the publishing ecosystem? In this study, we investigate patterns of citation behavior in astronomy to assess the presence and extent of such biases.

To explore these dynamics, we assembled a large bibliographic dataset of refereed astronomy publications spanning the years 2000 to 2025. We retrieved records from the SAO/NASA Astrophysics Data System (ADS) using a set of thematic keywords representative of diverse subfields in astronomy: \textit{Exoplanet, Star, Galaxy,} and \textit{AGN}. Our final dataset consists of 255,008 refereed articles.

These articles were published across 769 unique journals. However, approximately 62\% of them appear in just three major astronomy journals: \textit{The Astrophysical Journal} (ApJ), \textit{Monthly Notices of the Royal Astronomical Society} (MNRAS), and \textit{Astronomy \& Astrophysics} (A\&A). Note that \textit{The Astrophysical Journal} (ApJ) and \textit{ApJ Letters} are treated as a single journal in our analysis, as both  are indexed under the unified journal name "ApJ" in the ADS Bibcode records. \textit{The Astronomical Journal} (AJ) and \textit{The Astrophysical Journal Supplement Series} (ApJS) are listed separately and are not merged.

The fourth most frequent journal in our dataset is \textit{Physical Review D} (PhRvD), which contains approximately 3.3 times fewer articles than A\&A. Unlike the top three journals, PhRvD is primarily focused on theoretical astrophysics and cosmology and does not cover the full breadth of astronomical research. This thematic specialization may influence citation dynamics, as niche journals are less likely to be broadly cited by more generalist publications.

\begin{figure}[htbp]
    \centering
    \includegraphics[width=0.95\textwidth]{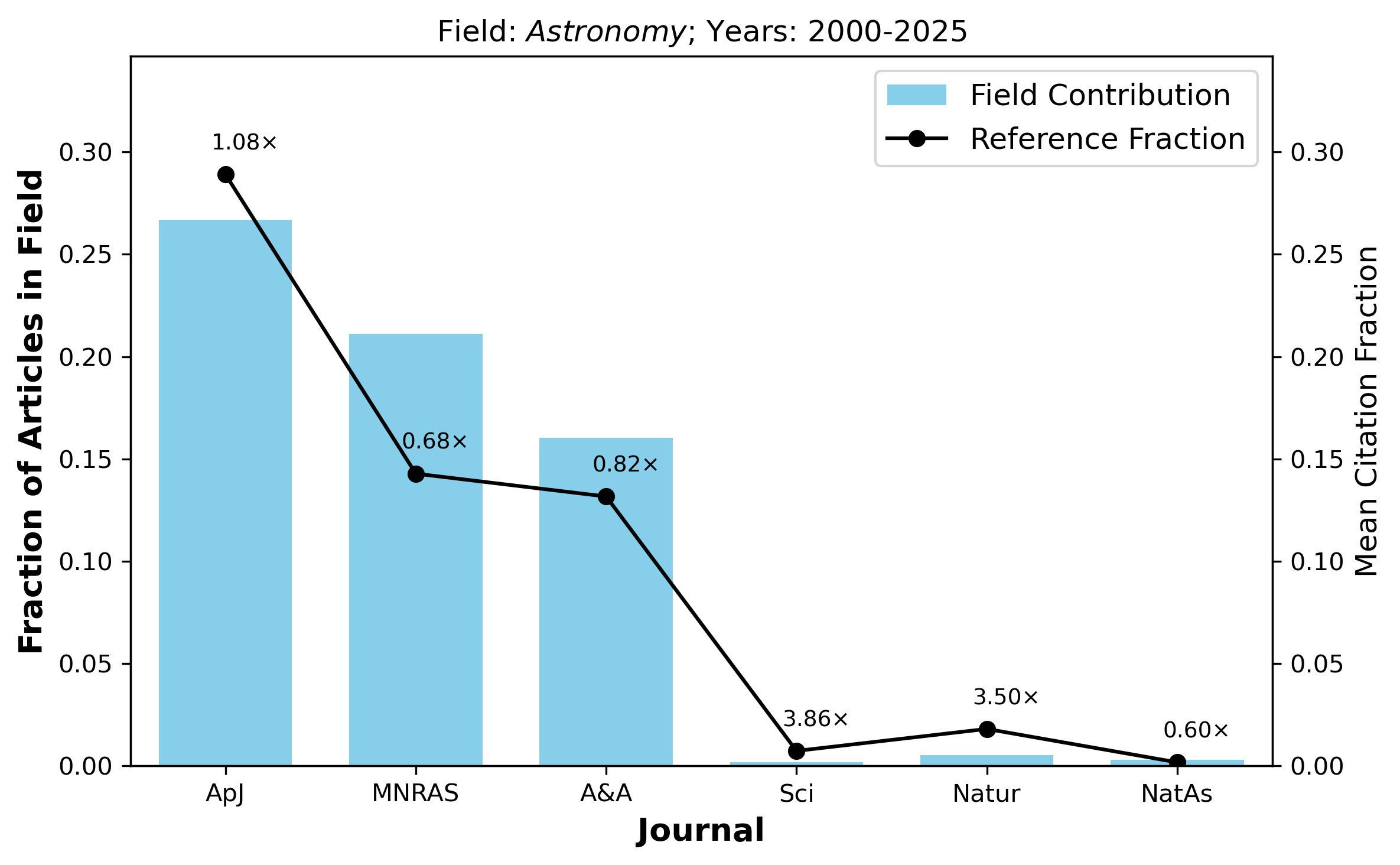}
    \caption{
Relative contribution of each journal to the total number of articles in the field (bar chart, left axis), compared with the mean fraction of references to that journal across all articles in the dataset (black circles, right axis). The numbers above the markers indicate the ratio between the mean reference fraction and the journal’s share of publications. This ratio highlights how frequently journals are cited relative to their publication volume.
}
    \label{fig:journal_contributions_bias_ALL_Astronomy}
\end{figure}

In this study, we concentrated on analyzing citation trends and potential biases in the three dominant astronomy journals: \textit{The Astrophysical Journal} (ApJ), \textit{Monthly Notices of the Royal Astronomical Society} (MNRAS), and \textit{Astronomy \& Astrophysics} (A\&A). We also included the relatively recent, high-impact journal \textit{Nature Astronomy} (NatAs), along with two broad-scope multidisciplinary journals, \textit{Nature} (Natur) and \textit{Science} (Sci). While our primary focus is on articles published in these six journals, all citations and references are analyzed using the full dataset of 255 thousand articles, allowing us to capture how each journal is cited across the broader astronomy literature. The share of articles published in the selected journals is illustrated in Figure~\ref{fig:journal_contributions_bias_ALL_Astronomy}. ApJ has the largest share, accounting for approximately 26\% of all articles, while NatAs and Sci have the smallest shares, each contributing around 0.25\%.

These publication shares have shifted over time  (see Methods). In the last 5 years, MNRAS has overtaken ApJ, contributing roughly 23\% of all articles, slightly more than ApJ. Meanwhile, \textit{Science} continues to have the smallest share of publications in our astronomy dataset (about 0.15\%), which is approximately three times fewer than \textit{Nature} and four times fewer than \textit{Nature Astronomy}.

To better understand the role and perceived impact of these journals within the field, we compared their share of publications with the average fraction of references they receive across all articles. This comparison is shown in Figure~\ref{fig:journal_contributions_bias_ALL_Astronomy}. A citation ratio greater than 1 indicates that a journal is referenced more frequently than would be proportional to its publication share, suggesting a relatively higher visibility or broader relevance. Conversely, a ratio below 1 suggests that a journal is cited less often relative to how much it contributes in terms of article volume.

For example, although \textit{ApJ} accounts for approximately 26\% of all articles, it is referenced slightly more often than its publication share (citation ratio $\sim$1.08). As noted earlier, we treat ApJ and \textit{ApJ Letters} as a single journal in this analysis, although the latter has a substantially higher impact factor\cite{Rithvik-25}. In contrast, \textit{MNRAS}, despite being the second most prolific journal, shows a noticeably lower citation ratio (around 0.68), indicating that it is cited less frequently relative to its publication volume. \textit{A\&A} shows a moderate citation deficit with a ratio of about 0.82.

On the other end of the spectrum, multidisciplinary journals such as \textit{Science} and \textit{Nature} receive citation fractions far exceeding their publication shares, with citation ratios of 3.86 and 3.50, respectively. This aligns with their longstanding reputation for high-impact publications and their broader readership across disciplines, despite the relatively small number of astronomy-related articles they publish. These journals, while classified as multidisciplinary, have been shown to consistently overrepresent certain high-visibility fields \cite{Redondo-24}, which may further inflate their citation ratios within specific topics. Moreover, journal venue itself exerts a causal influence on citation counts, independent of article quality \cite{Traag-21}, reinforcing the role of journal prestige in amplifying scientific visibility. This self-reinforcing dynamic—where already prominent journals accumulate more citations because of their visibility rather than the intrinsic merit of individual articles—is consistent with the well-known "Matthew Effect" in science \cite{Merton-68}. It reflects how prestige and early recognition can lead to cumulative advantages in scholarly attention.

Interestingly, \textit{Nature Astronomy}, a relatively new and specialized journal, currently receives slightly fewer citations relative to its share of publications (ratio $\sim$0.60). However, this has improved in recent years, reaching a ratio of 0.80 in the most recent 5-year period. Similar upward trends are seen for both \textit{Science} and \textit{Nature}, indicating increasing citation intensity for articles published in these high-profile journals (see Methods).

\begin{figure}[ht]
    \centering
    \includegraphics[width=0.95\linewidth]{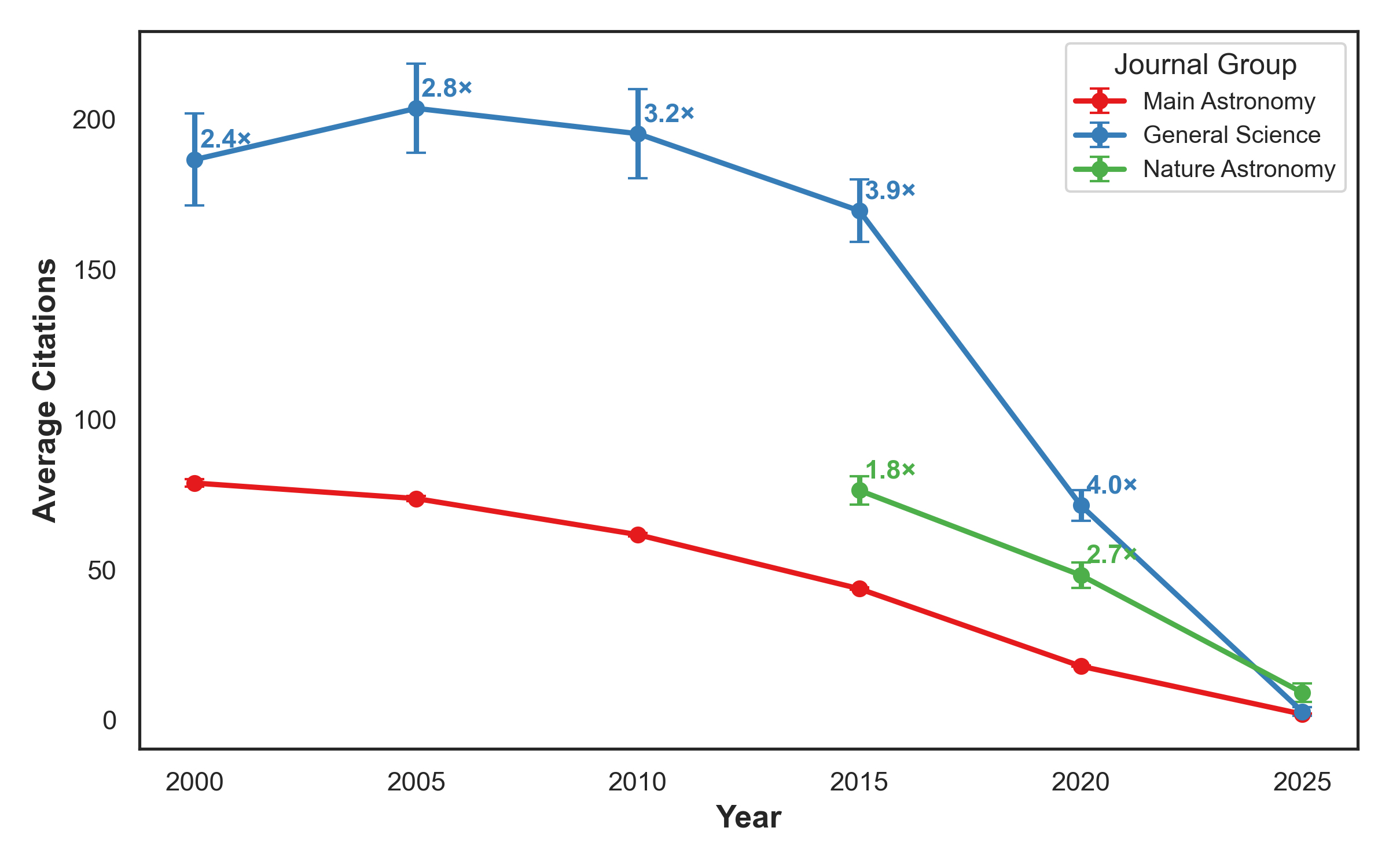}
    \caption{
        Average citation counts per 5-year bin for three groups of journals in the \textit{Astronomy} field.
        The groups include major astronomy journals (\textit{A\&A}, \textit{ApJ}, \textit{MNRAS}), general science journals (\textit{Science}, \textit{Nature}), and \textit{Nature Astronomy}.
        Error bars represent the standard error of the mean citation count per bin. 
        Ratios indicate how much higher the average citation count in each bin is for general science or \textit{Nature Astronomy} compared to the main astronomy journals.
    }
    \label{fig:avg_citations_grouped}
\end{figure}

The impact of the journals can also be assessed by comparing the average number of citations their articles receive over time. As shown in Figure~\ref{fig:avg_citations_grouped}, general science journals (\textit{Nature} and \textit{Science}) consistently attract significantly more citations than the main astronomy journals (ApJ, A\&A, and MNRAS). Notably, the citation advantage of general science journals is more pronounced for more recently published articles. The citation ratio between general science and astronomy journals increases from approximately 2.5 for articles published between 2000--2004 to about 4.0 for those published between 2020--2024. This trend might suggest that general science journals tend to publish articles on high-visibility or trending topics that accumulate citations rapidly\cite{Traag-21}, particularly within the first decade after publication. Other factors, such as the number of references included in each article—which has been shown to correlate positively with citation count—may also contribute to these differences. Editorial norms and field-specific practices often lead to shorter reference lists in multidisciplinary journals, particularly in the past, which may have historically amplified in-journal citation ratios \cite{Mattiazzi-24}.

Interestingly, it has been shown that highly ranked journals are predominantly governed by preferential attachment, where references are selected more carefully and with intent, allowing already visible venues to accumulate further attention, while citations to lower-ranked journals tend to follow a more uniform, accidental distribution \cite{Mrowinski-22}.

\begin{figure}[htbp]
    \centering
    \includegraphics[width=0.95\textwidth]{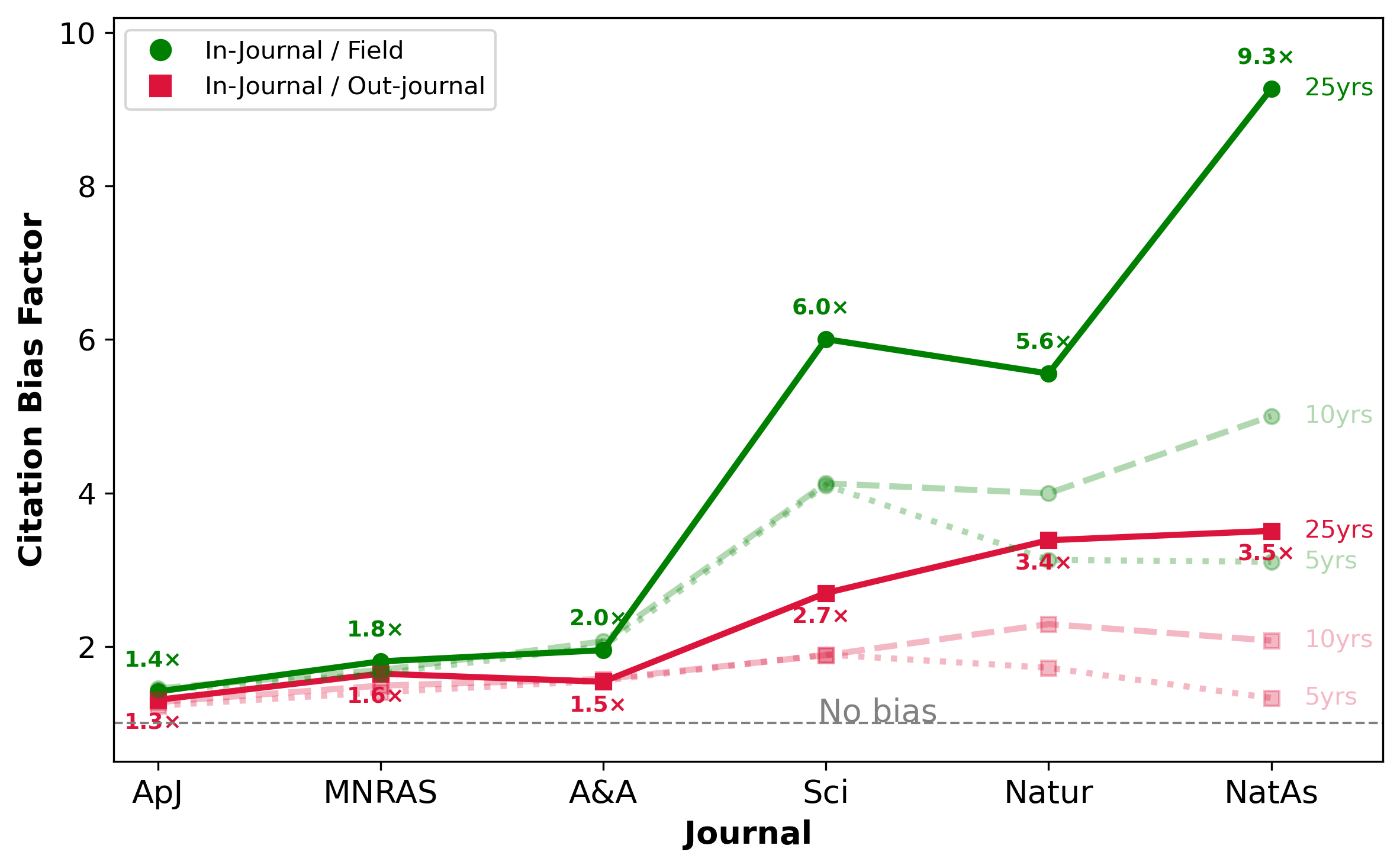}
    \caption{
        Citation bias ratios across journals. The green lines show the ratio between the mean percentage of references to a journal when an article is published \textit{in that journal}, and the mean percentage of references to that journal across the entire field (\textit{In-Journal / Field}). The red lines display the ratio between the mean percentage of references to a journal when an article is published \textit{in that journal}, versus when the same first author publishes in other journals (\textit{In-Journal / Outside-Journal}). Solid, dashed, and dotted lines correspond to results over the full 25-year period, the last 10 years, and the last 5 years, respectively. A horizontal gray dashed line at 1.0 indicates no citation bias.
    }
    \label{fig:journal_bias_curves_Astronomy_name}
\end{figure}

Figure~\ref{fig:journal_bias_curves_Astronomy_name} illustrates citation bias metrics for a selection of major journals. The figure contrasts how frequently authors cite a given journal when they publish within it, compared to either the field-wide citation average (green line) or their own citation behavior when publishing in other journals (red line). Specifically, the red line represents the ratio of the average fraction of references to a journal when the first author publishes in that journal, compared to when the same first author publishes elsewhere as a firstauthor. Values above 1 indicate a tendency to cite a journal more frequently when publishing in it, reflecting varying degrees of preferential citation or venue-linked citation behavior.

The results show that all journals exhibit some level of in-journal citation bias, as reflected by the green line. However, the magnitude of this bias remains modest for field-specific journals such as ApJ, MNRAS, and A\&A, where the In-Journal / Field ratios stay below 2. This may partly reflect the fact that these journals are associated with distinct regional communities—ApJ with the United States, MNRAS with the United Kingdom, and A\&A with continental Europe—which tend to attract submissions and citations from their respective scientific networks. This likely reflects both institutional publishing preferences and natural topical clustering, where subfields tend to coalesce around specific journals, leading to more intra-journal citations without intentional bias. Such clustering likely explains some of the elevated ratios for multidisciplinary journals as well. Since journals like \textit{Science} and \textit{Nature} are highly selective and do not cover all research topics within astronomy, the articles they publish are often on prominent, high-visibility themes. This topical filtering can amplify the concentration of citations within the same journal. Moreover, citation patterns may also be shaped by institutional or group-based publishing practices. It is common for research groups (sometimes entire institutions or countries) to develop journal-specific submission habits. When group members cite previous work from their team, which is often published in the same journal, it can contribute to an elevated In-Journal / Field ratio. 

The red curves, representing the In-Journal / Out-Journal comparison for the same first authors, are typically lower than the green ones, suggesting that some portion of the citation clustering may be linked to author identity rather than topical focus alone.

However, this explanation falls short in the case of general science journals such as \textit{Science} and \textit{Nature}, where the In-Journal / Field ratios reach extreme values (up to 9.3), while the red line shows a more moderate increase. It is unlikely that individual authors or research groups consistently publish across multiple articles in these highly selective journals. Instead, these patterns may reflect broader behavioral dynamics, such as a tendency for authors to cite these journals strategically, potentially to appeal to editors, underscore the relevance of the topic, or demonstrate that similar high-impact work has previously appeared in the same journal. Such motivations, while not necessarily deliberate bias, can nonetheless shape citation behavior in ways that reinforce the visibility and perceived prestige of these journals.

It is worth noting, however, that despite the elevated in-journal citation rates, the contribution of these citations to the impact factors of multidisciplinary journals such as \textit{Science} and \textit{Nature} is likely negligible. These journals publish relatively few astronomy-related articles, and their impact factors are driven by a large volume of citations across a wide range of disciplines. Thus, while strategic or habitual self-citation may occur at the level of individual papers or authors, it does not inflate journal-level metrics in these cases.

The dashed and dotted lines in Figure~\ref{fig:journal_bias_curves_Astronomy_name} show how these citation bias ratios have evolved over time, specifically for the last 10 years (dashed) and last 5 years (dotted). A clear downward trend is visible across most journals, particularly for the multidisciplinary ones. While the 25-year curves show high In-Journal / Field ratios for \textit{Science} and \textit{Nature} (above 6 and even 9), these values drop significantly in recent years, approaching the levels seen in the main astronomy journals. The observed decline in citation bias ratios may reflect a combination of factors, including greater awareness of citation ethics, broader collaboration networks, and more standardized review processes across journals. Additionally, the widespread adoption of preprint platforms such as arXiv has improved access to scientific articles regardless of journal venue, potentially reducing the influence of journal prestige on citation behavior. As a result, citation practices in multidisciplinary journals have become more aligned with those of field-specific journals, at least over the last decade.

Altogether, these findings underscore the complex interplay between editorial policies, publication practices, and citation dynamics. While some level of in-journal citation bias is natural and even expected, its magnitude and evolution can offer insight into the broader forces shaping how scientific recognition is distributed across journals. The persistent disparities we observe, particularly the strong citation inflation in multidisciplinary journals, highlight the need for greater awareness of how citation patterns are influenced by structural and behavioral factors, rather than purely scientific merit.

To promote a more equitable and transparent research ecosystem, we encourage all participants in the publication process-authors, reviewers, and editors alike—to reflect critically on citation choices. References should be selected based on relevance and contribution to the scientific context, not solely on journal prestige, familiarity. Increased attentiveness to fair citation practices can help ensure that scientific credit is distributed more justly, and that valuable work from all corners of the community is appropriately acknowledged.

These insights could inform future discussions on research evaluation practices, especially in contexts where citation-based indicators are used to assess scientific impact or allocate funding.


\clearpage
\newpage
\section*{Methods}

\subsection*{Data Retrieval from the ADS API}
\label{ads_api}

To analyze trends in astronomy-related research, we retrieved bibliographic data from the SAO/NASA Astrophysics Data System (ADS) using its public API. Our focus was on refereed publications between 2000 and 2025 that included one or more of the following keywords: \textit{Exoplanet, Star, Galaxy,} and \textit{AGN}. These keywords are widely used across the astronomy community and span a broad range of scientific topics. Importantly, articles with these labels are typically published in the main astronomy journals—\textit{ApJ}, \textit{MNRAS}, and \textit{A\&A}—ensuring thematic consistency and comparability across venues.

We chose this keyword-based approach instead of using the entire ADS 'astronomy' database, which includes a large number of articles focused on geophysics and theoretical cosmology. Such articles frequently appear in journals like \textit{GeoRL} (geosciences) or \textit{PhRvD} (cosmology and high-energy physics), which are outside the scope of the main astronomy journals and are rarely cited in or by them. Including these topics would introduce multidisciplinary citation behavior that could obscure the journal-level patterns our study aims to evaluate. Nonetheless, we verified that our main findings and conclusions remain robust even when the broader dataset is considered.

For each keyword, we queried all refereed journal articles and extracted metadata including publication venue, year, citations, abstracts, author affiliations, reference lists, and Digital Object Identifiers (DOIs).

It is worth noting that, in some cases, articles indexed in ADS may lack complete metadata—such as missing author lists or reference entries—due to variations in journal submission standards or ingestion issues during indexing. Although such cases are relatively rare, they can affect aggregate counts. Our filtering therefore ensures that only articles with the minimum required metadata—at least one author and one or more references—are included in the final dataset.

A summary of article counts per keyword is provided in Table~\ref{tab:keyword_counts}. In total, we retrieved 373,531 records, of which 269,145 were unique by DOI. Among these, 268,757 included at least one listed author. After removing 13,749 articles without reference lists, our final dataset comprised 255,008 unique, refereed journal articles with complete authorship and citation metadata.

\begin{table}[htbp]
\centering
\caption{Number of articles retrieved from ADS by keyword (2000--2025).}
\label{tab:keyword_counts}
\begin{tabular}{lr}
\toprule
\textbf{Keyword} & \textbf{Articles} \\
\midrule
AGN & 18,651 \\
Exoplanet & 12,571 \\
Galaxy & 154,130 \\
Star & 185,370 \\
\midrule
\textbf{Total retrieved} & \textbf{373,531} \\
\textbf{Total unique} & \textbf{269,145} \\
\textbf{With author metadata} & \textbf{268,757} \\
\textbf{With at least one reference} & \textbf{255,008} \\
\bottomrule
\end{tabular}
\end{table}

\subsection*{Extracting Journal names and References}
\label{affiliation_country}

To determine the journal in which each article was published, we extracted the primary journal name from the \texttt{bibstem} field, which is provided by the ADS database as a stringified list of potential journal abbreviations. We parsed this list and selected the first entry as the primary journal. 

To identify the journals cited in each article, we extracted journal codes from the ADS-style reference strings using a custom parser. Each reference is represented by a bibcode, from which we isolated the journal code by capturing the character sequence between the 4-digit publication year and the first dot. For instance, the bibcode \texttt{1998PhRvL..82..896Z} was parsed to extract \texttt{PhRvL} as the journal code. This allowed us to compile a list of referenced journals for every article in our dataset.

\subsection*{First Author Affiliation and Country}
\label{affiliation_country}

\begin{figure}[htbp]
    \centering
    \includegraphics[width=0.95\textwidth]{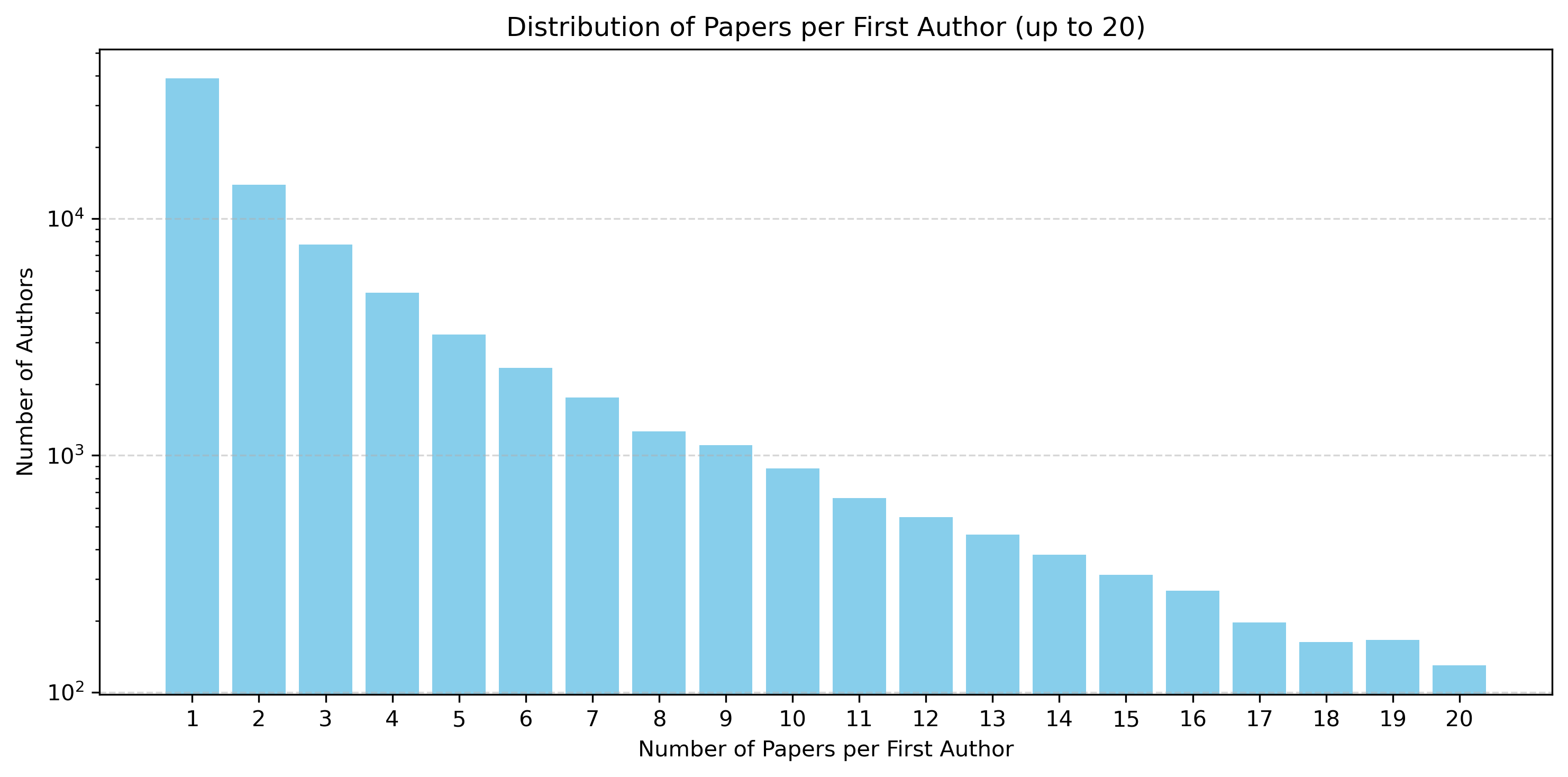}
    \caption{
        Histogram showing the number of first authors with a given number of publications (from 1 to 20).
        The data are based on unique author identifiers. 
        Most authors appear only once or twice, suggesting a high number of occasional contributors or partially name mismatches.
    }
    \label{fig:author_pub_hist}
\end{figure}

In addition to retrieving bibliographic and citation data, we also compiled basic authorship and journal-level statistics from the ADS records. Based on the journal names as listed in ADS, we identified 769 unique journals represented in our final sample. Similarly, using the names of the first authors as they appear in ADS for each article, we identified 80,304 unique first authors. Figure~\ref{fig:author_pub_hist} shows a histogram of the number of first authors with a given number of publications (from 1 to 20). 

It is important to note that the number of unique first authors may not reflect the true number of individuals, due to variations in name formatting across different publications and journals. Authors may appear under slightly different names depending on journal conventions, transliterations, or personal choices such as using initials, full names, or even changing names over time. For example, the first author of this manuscript (Adibekyan) has 25 first-author articles in our dataset, but they appear under three distinct forms: \textit{Adibekyan, V. Zh.} (12 articles), \textit{Adibekyan, V.} (10 articles), and \textit{Adibekyan, Vardan} (3 articles).

On the other hand, name similarity across different individuals may result in an overcount of publication counts for a given name. This effect could in principle be mitigated by considering first-author affiliations; however, institutional affiliations are often inconsistently formatted and researchers frequently change institutions over the course of their careers. As a result, disambiguating author identities based on affiliation data is highly challenging and was not attempted in this study. Alternatively, one could identify the country of the affiliation and use it for crossmatching the authors. 

We extracted and standardized country information from the first author's affiliation string. These fields are often inconsistently formatted and include a mix of institutional names, postal codes, abbreviations, and other location descriptors, making automated extraction non-trivial.

Our approach was implemented in a custom Python class, \texttt{AdsAffCountry}, which followed a multi-step procedure. Initially, we parsed the raw affiliation---usually a stringified list---and extracted the final comma-separated segment of the first listed affiliation, assuming it most likely contained a geographic identifier. However, this final segment frequently included ambiguous elements such as U.S. state codes or postal zones. We applied heuristic mappings to convert these to their corresponding countries, including numerous variants of ``USA''.

Next, we normalized affiliation suffixes using the \texttt{country\_converter} library, mapping free-form text to standardized country names. This step resolved many cases but still left unresolved entries due to institution names, acronyms, or non-standard formats (e.g., ``ESO'', ``Observatoire de Paris''). We then compiled the 200 most frequent unresolved segments and manually assigned countries using contextual knowledge, institutional naming conventions, and support from ChatGPT.

Validation against a curated list of country names was applied throughout, and the final dataset with standardized first author country labels was saved for downstream statistical analyses.

Of the 268,757 unique articles with authors, 12,279 (4.6\%) lacked any first author affiliation string from which a country could be extracted. A simple suffix-based method yielded no valid country for 95,593 articles (35.6\%). After the second iteration, this was reduced to 58,779 (21.9\%), recovering 36,814 entries (13.7\%). A final iteration of normalization and manual mapping further decreased the unresolved set to 16,950 (6.3\%), recovering an additional 41,829 entries (15.6\%).

In total, we successfully recovered country labels for 78,643 articles beyond the initial extraction, yielding a final total of 251,807 articles (93.7\%) with valid first author country information.

\begin{figure}[htbp]
    \centering
    \includegraphics[width=0.75\textwidth]{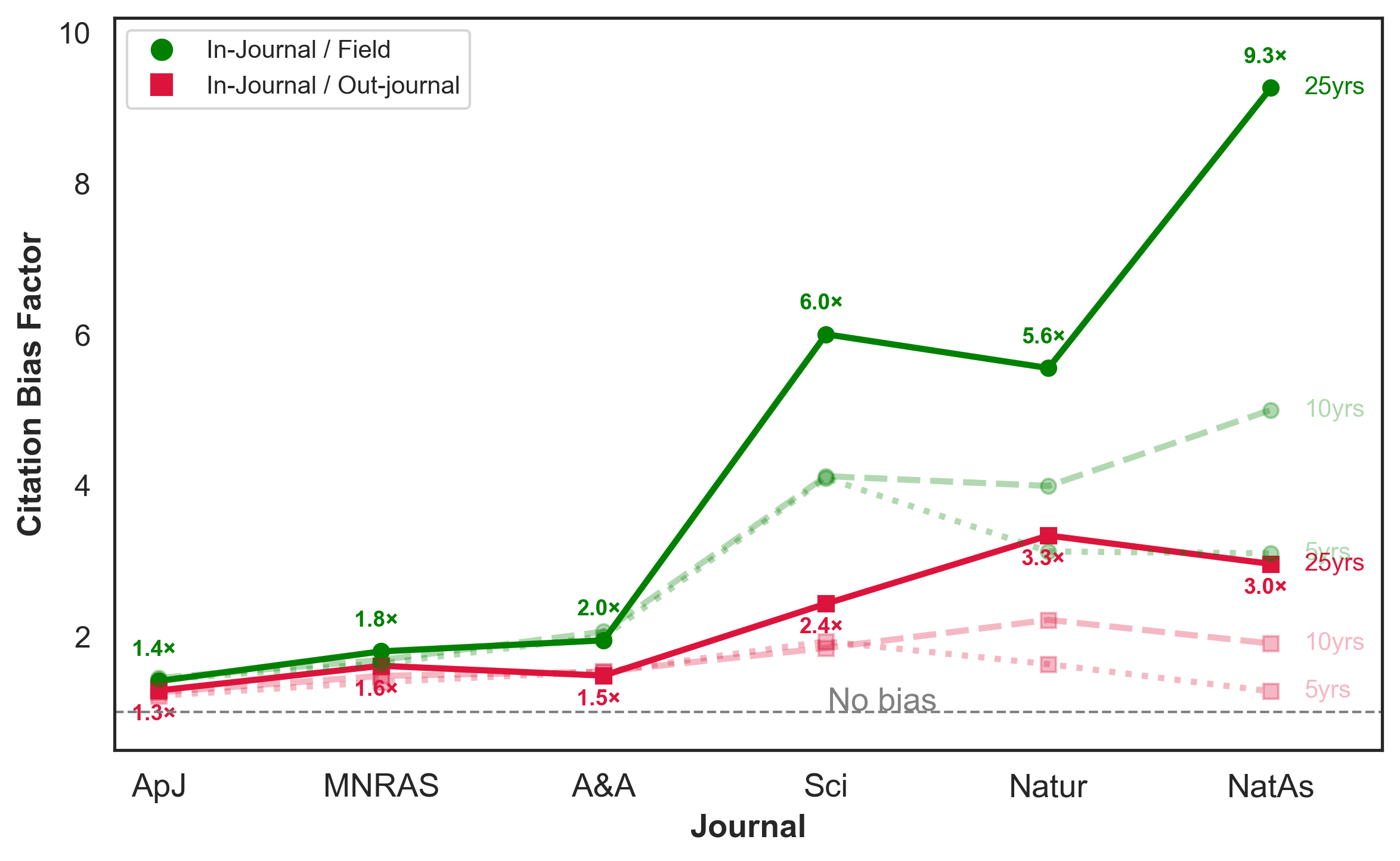}
    \caption{
        Same as Figure~\ref{fig:journal_bias_curves_Astronomy_name}, except that first-author crossmatching is based on both author names and the countries of their affiliations.
    }
    \label{fig:journal_bias_curves_Astronomy_name_country}
\end{figure}

Figure~\ref{fig:journal_bias_curves_Astronomy_name_country} presents the same analysis as in Figure~\ref{fig:journal_bias_curves_Astronomy_name}, but with first-author crossmatching performed using both author names and the countries of their affiliations. Overall, the results remain consistent, although the In-Journal / Outside-Journal citation bias is slightly reduced for multidisciplinary journals.

\subsection*{Temporal Trends in Journal Representation and Citation Share}
\label{sec:journal_trends}

To investigate how journal representation and citation impact evolve over time, we analyzed the relative publication share and mean reference fraction of selected journals for three distinct time periods: 2000--2025, 2015--2025, and 2020--2025. Table~\ref{tab:journal_counts_percentages} summarizes the number and percentage of articles published in each journal across these intervals.

\begin{figure}[htbp]
    \centering
    \begin{subfigure}[b]{0.65\textwidth}
        \centering
        \includegraphics[width=\textwidth]{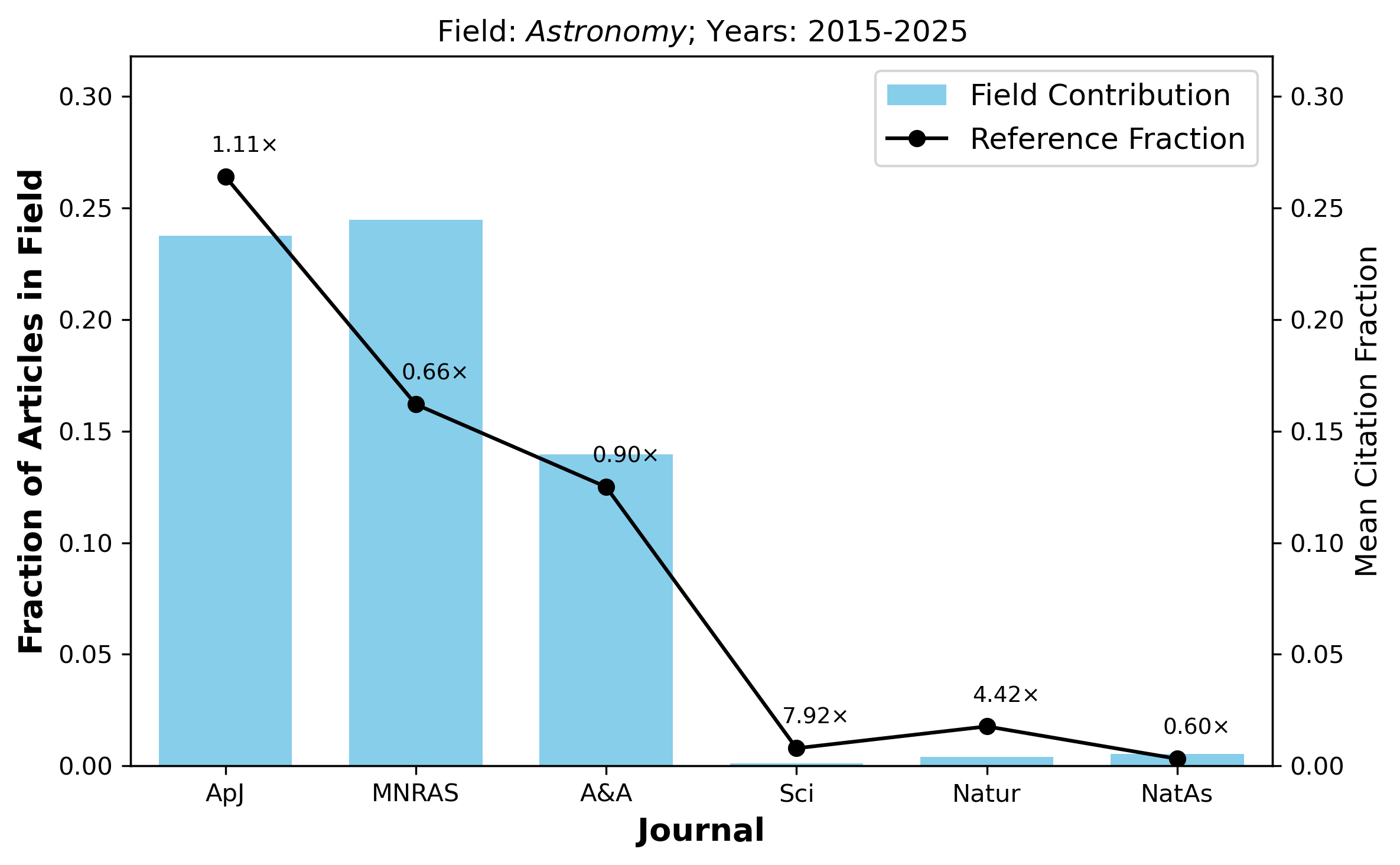}
    \end{subfigure}
    
    \vspace{-0.3cm}
    
    \begin{subfigure}[b]{0.65\textwidth}
        \centering
        \includegraphics[width=\textwidth]{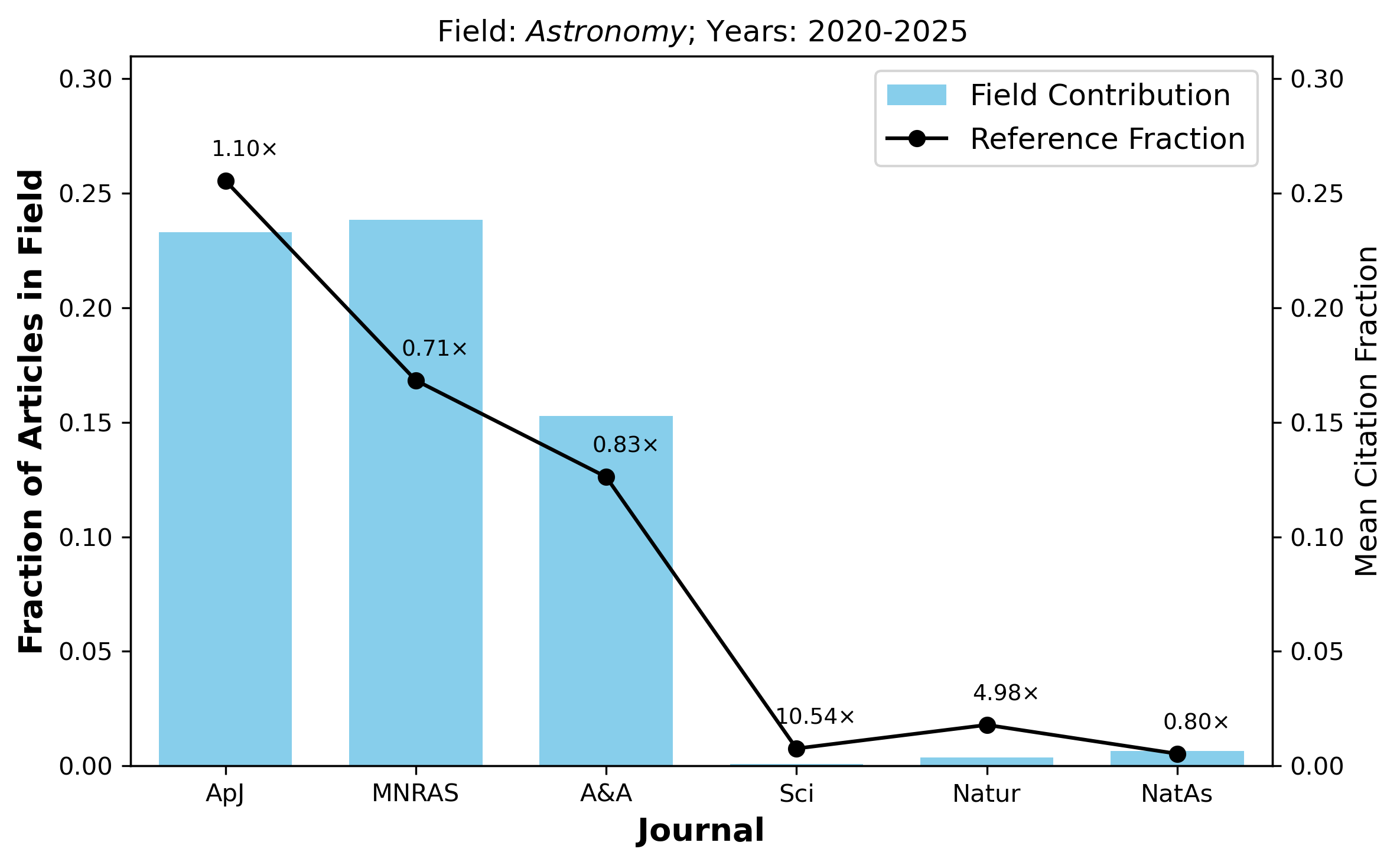}
    \end{subfigure}
    \caption{
Same as Figure~\ref{fig:journal_contributions_bias_ALL_Astronomy}, but for the last 10 years (top) and last 5 years (bottom).
    }
    \label{fig:journal_contributions_bias_recent}
\end{figure}

We computed the citation bias ratio for each journal, defined as the ratio between the average fraction of references to that journal and its share of published articles in the dataset. This ratio reflects how frequently journals are cited relative to how much they contribute to the field’s publication volume. Figure~\ref{fig:journal_contributions_bias_ALL_Astronomy} shows the results for the full 2000--2025 range, while Figure~\ref{fig:journal_contributions_bias_recent} presents the same analysis for the last 10 years (top panel) and the last 5 years (bottom panel).

The results show that multidisciplinary journals tend to be cited at rates much higher than their share of published articles. Although \textit{Science} and \textit{Nature} together account for less than 1\% of astronomy articles, they receive citation fractions 5 to 10 times greater than their publication share. This pattern is most prominent in the most recent 5-year window, where the citation ratio for \textit{Science} exceeds 10. 

In contrast, MNRAS and A\&A are cited at lower rates relative to their publication volume, with ratios around 0.7--0.8. \textit{The Astrophysical Journal} (ApJ) shows more balanced citation behavior, with ratios close to or slightly above 1. \textit{Nature Astronomy}, despite being a newer journal with high visibility, currently shows citation ratios below 1 in both the 10- and 5-year windows.

These findings may suggest that citation behavior is shaped not only by article content and visibility, but also by broader factors such as audience reach, topic selection, and editorial practices. Journals with broader scope or higher perceived impact tend to be cited more frequently relative to their contribution to the overall publication landscape, while specialized field journals, despite their essential role, tend to receive proportionally fewer citations.

\begin{table}[htbp]
\centering
\caption{Publication counts and relative percentages for selected journals over three time periods: 2000--2025, 2015--2025, and 2020--2025.}
\label{tab:journal_counts_percentages}
\begin{tabular}{llrr}
\toprule
\textbf{Period} & \textbf{Journal} & \textbf{Count} & \textbf{Percentage (\%)} \\
\midrule
\multirow{6}{*}{2000--2025}
    & ApJ     & 65,402 & 25.6 \\
    & MNRAS   & 52,887 & 20.7 \\
    & A\&A    & 39,208 & 15.4 \\
    & Sci     &    625 & 0.25 \\
    & Natur   &  1,553 & 0.61 \\
    & NatAs   &    698 & 0.27 \\
\midrule
\multirow{6}{*}{2015--2025}
    & ApJ     & 30,997 & 23.0 \\
    & MNRAS   & 31,917 & 23.6 \\
    & A\&A    & 18,511 & 13.7 \\
    & Sci     &    221 & 0.16 \\
    & Natur   &    644 & 0.48 \\
    & NatAs   &    698 & 0.52 \\
\midrule
\multirow{6}{*}{2020--2025}
    & ApJ     & 15,699 & 22.5 \\
    & MNRAS   & 16,017 & 23.0 \\
    & A\&A    & 10,432 & 15.0 \\
    & Sci     &    106 & 0.15 \\
    & Natur   &    312 & 0.45 \\
    & NatAs   &    446 & 0.64 \\
\bottomrule
\end{tabular}
\end{table}

\subsection*{Citations and References}
\label{cite_ref}

\begin{figure}[htbp]
    \centering
    \includegraphics[width=0.85\textwidth]{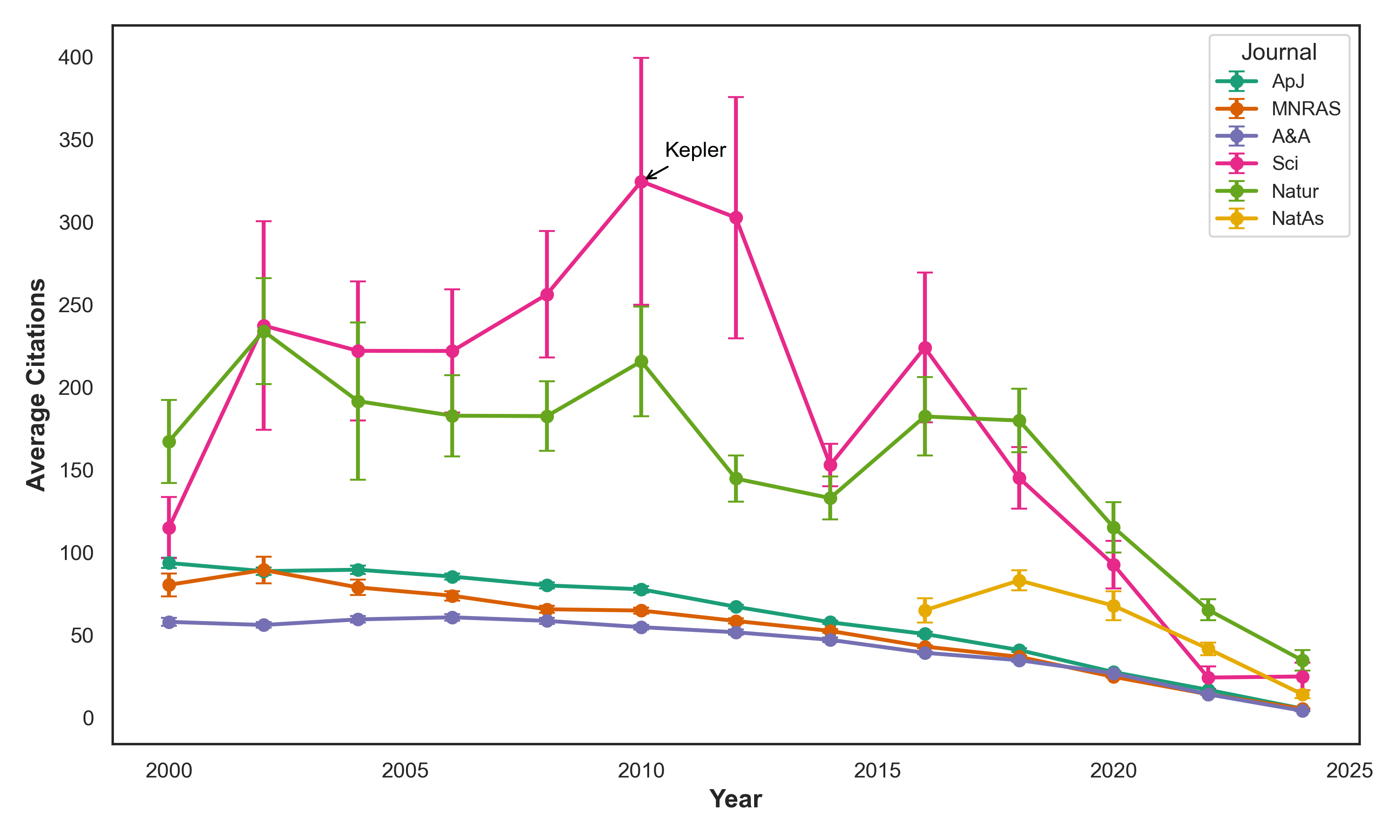}
    \caption{
        \textbf{Average citations per 2-year bin for selected astronomy journals.}
         For each journal, the figure shows the mean number of citations (with standard error) received by articles published in that 2-year period. The error bars represent the standard error of the mean. The prominent citation peak around 2010 for \textit{Science} corresponds to the publication and early impact of major discoveries from the \textit{Kepler} mission. Other secondary peaks observed in \textit{Science} and \textit{Nature} are associated with individual high-impact discoveries.
    }
    \label{fig:avg_citations_by_year_bin_Astronomy_seaborn}
\end{figure}

In Figure~\ref{fig:avg_citations_by_year_bin_Astronomy_seaborn}, we show the average number of citations received by articles from selected journals as a function of publication year, grouped in 2-year bins. Each point represents the mean citation count for articles published in that period, with error bars denoting the standard error of the mean. As expected, older articles tend to accumulate more citations, although the citation trajectories vary notably across journals. In particular, \textit{Science} shows a prominent citation peak around 2010, corresponding to the publication and early impact of major discoveries from the \textit{Kepler} mission. Other secondary peaks in \textit{Science} and \textit{Nature} are also associated with individual high-impact discoveries, reflecting the strong citation response to landmark findings published in these multidisciplinary venues.

\begin{figure}[htbp]
    \centering
    \includegraphics[width=0.85\textwidth]{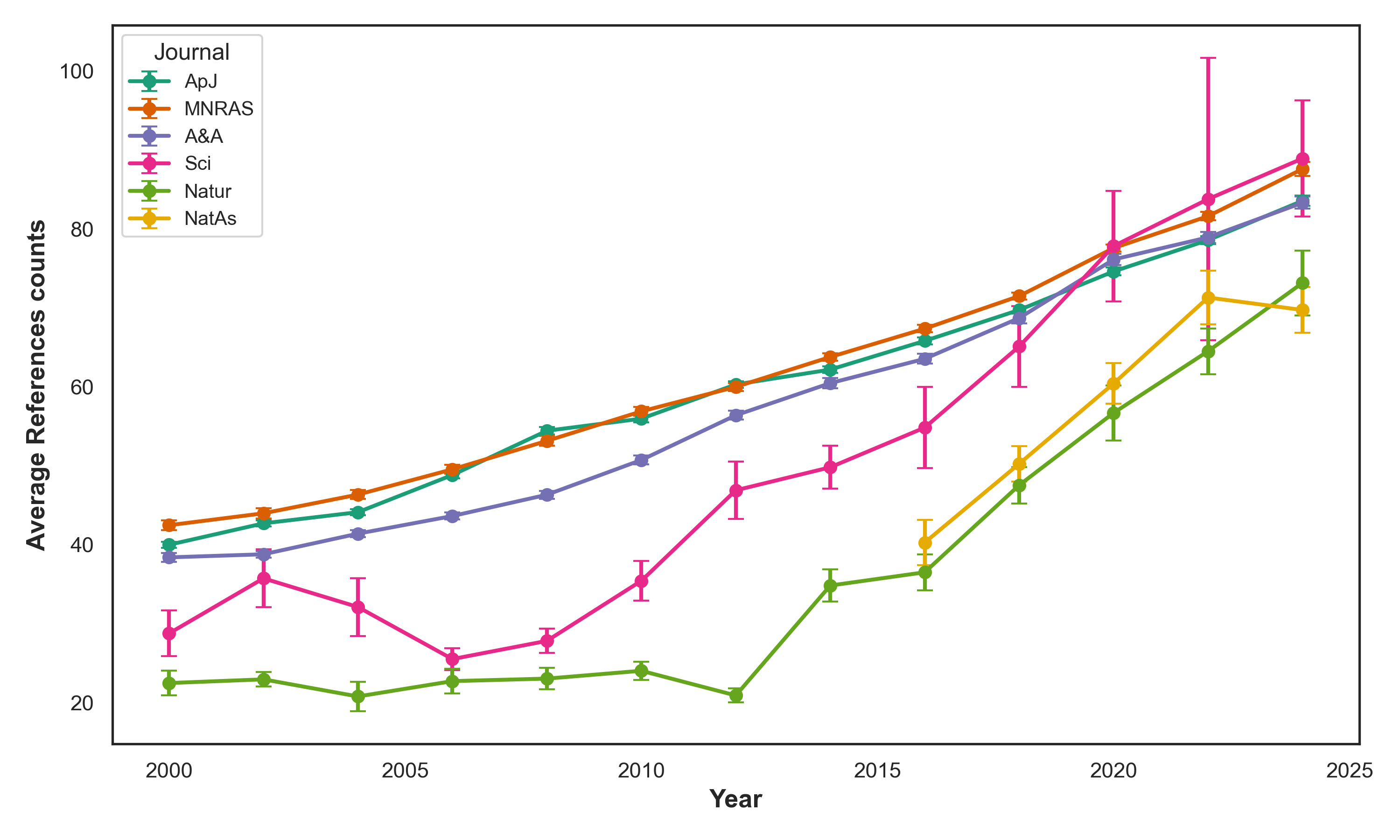}
    \caption{
        \textbf{Average counts of references per 2-year bin for selected astronomy journals.}
         For each journal, the figure shows the mean number of references (with standard error) in articles published in that 2-year period. The error bars represent the standard error of the mean. 
    }
    \label{fig:avg_references_by_year_bin_Astronomy_seaborn}
\end{figure}

\begin{figure}[htbp]
    \centering
    \includegraphics[width=0.85\textwidth]{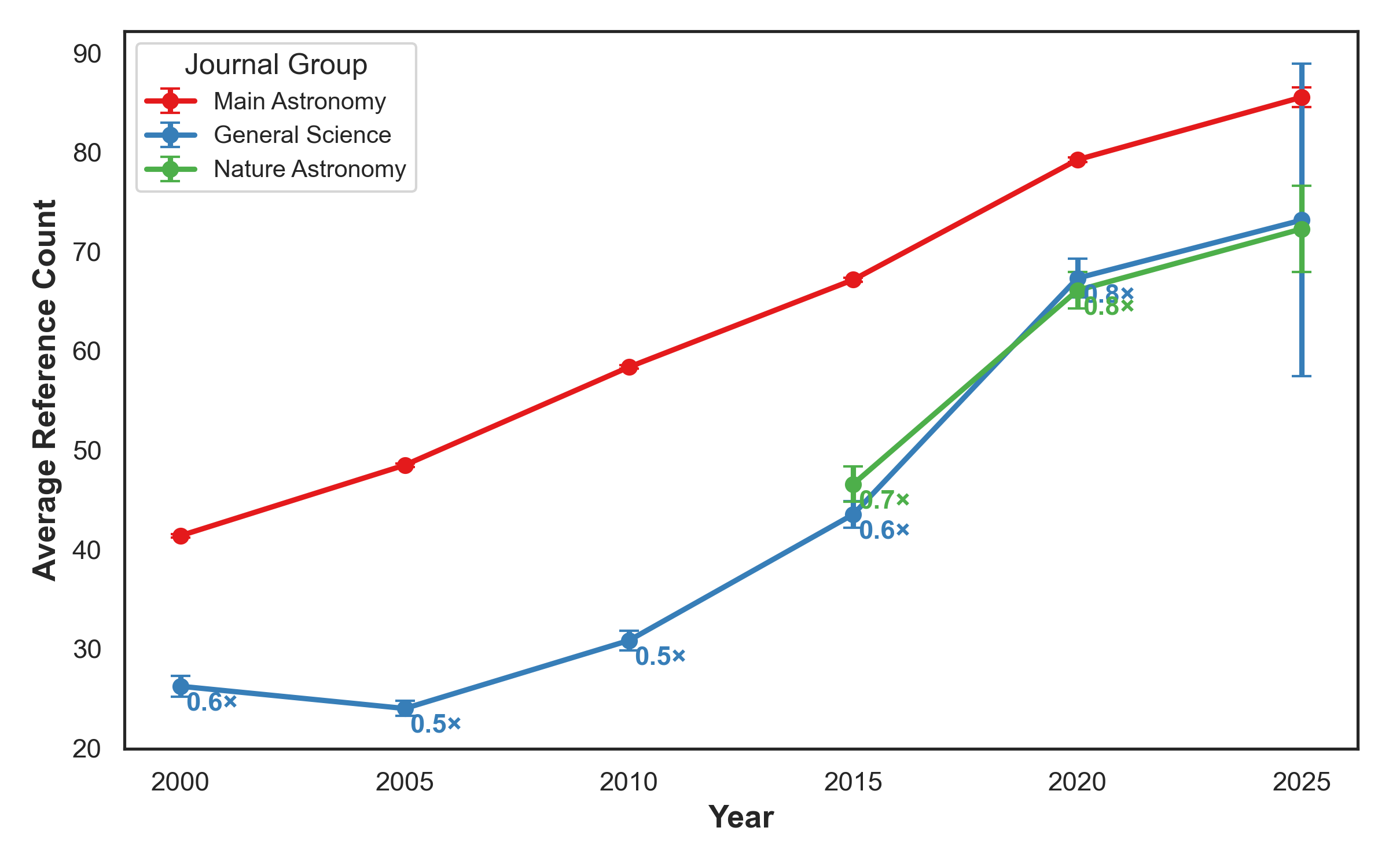}
    \caption{
        \textbf{Average reference counts per 5-year bin for three groups of journals in the \textit{Astronomy} field.}
         The groups include major astronomy journals (\textit{A\&A}, \textit{ApJ}, \textit{MNRAS}), general science journals (\textit{Science}, \textit{Nature}), and \textit{Nature Astronomy}.
        Error bars represent the standard error of the mean reference count per bin. 
        Ratios indicate how much lower the average reference count in each bin is for general science or \textit{Nature Astronomy} compared to the main astronomy journals. 
    }
    \label{fig:avg_references_grouped_with_ratios_Astronomy}
\end{figure}

As shown in Figures~\ref{fig:avg_references_by_year_bin_Astronomy_seaborn} and ~\ref{fig:avg_references_grouped_with_ratios_Astronomy}, the reference practices in general science journals have changed significantly over time. In the early 2000s, these journals included far fewer references per article than typical astronomy journals—around half as many on average. In recent years, however, their reference counts have increased substantially, approaching the levels found in specialized astronomy journals. This shift may be a response to broader multidisciplinary expectations or editorial policy changes encouraging more thorough citation practices.

\begin{figure}[htbp]
    \centering
    \includegraphics[width=0.85\textwidth]{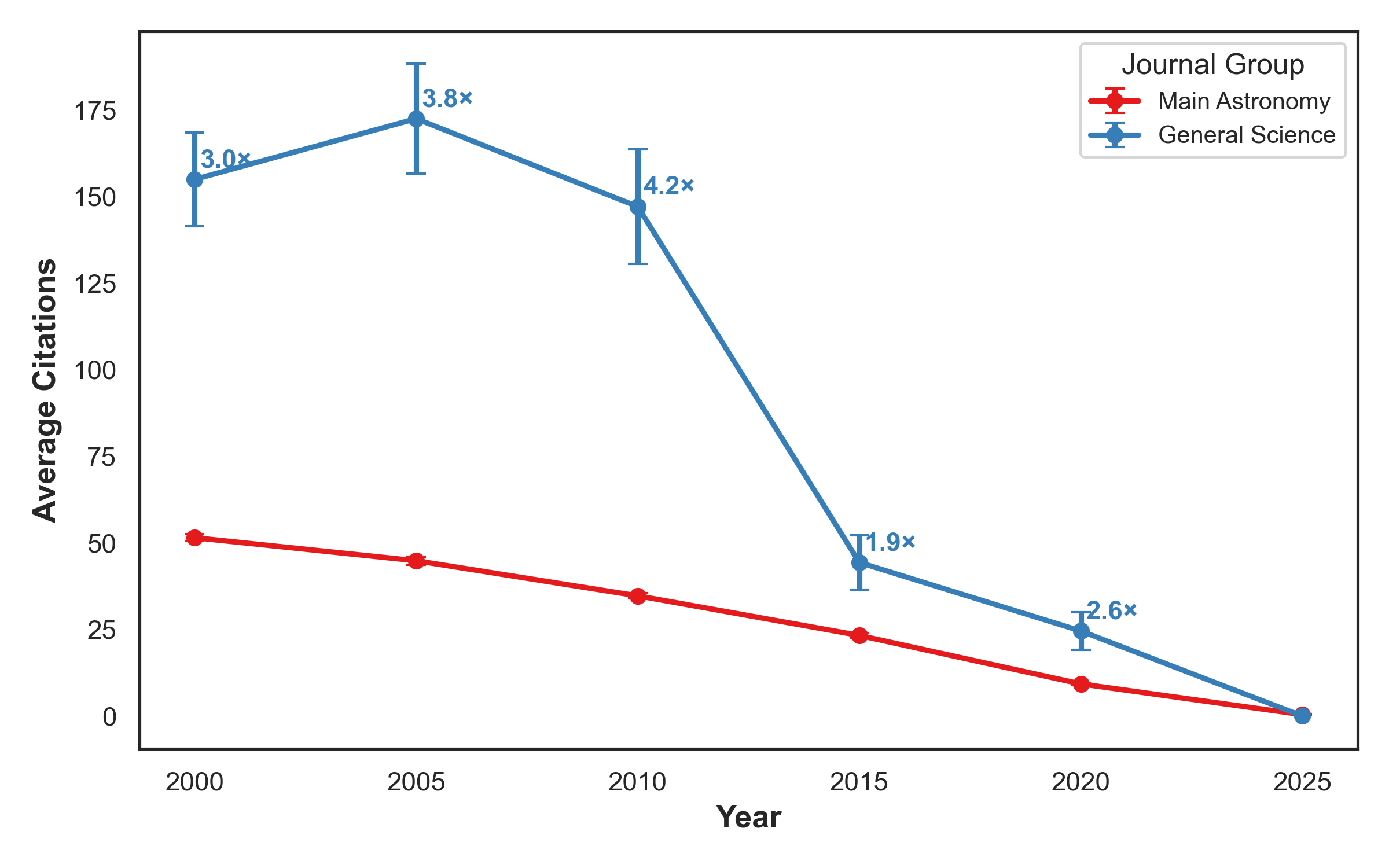}
    \caption{
          Same as Figure~\ref{fig:avg_citations_grouped}, except that only articles with up to 30 articles in the references are considered. 
    }
    \label{fig:avg_citations_grouped_with_ratios_Astronomy_30references}
\end{figure}

Reference counts correlates with the citation counts, however, this might not be the only reason of the higher citation rates of multidisciplinary journals during the recent years when compared to the old articles as when restricting the sample to only those with small reference numbers (up to 30 references maximum) the factor overcitations of articles in multidisciplinary journals during the recent years decreases when compared with the old articles (Figure~\ref{fig:avg_citations_grouped_with_ratios_Astronomy_30references}). At the same time, when focusing only on these articles the overcitation factor for the general science articles increases.

Together, these trends suggest that while general science journals benefit from exceptionally high visibility and citation rates, their historical 'undercitation' may have partially reflected more concise referencing norms. The rise in their reference counts coincides with a slightly reduced citation advantage, potentially indicating that broader referencing has helped normalize their impact relative to field-specific journals. These patterns highlight how citation metrics are influenced not only by journal scope and prestige but also by evolving disciplinary norms in referencing behavior.

\bibliography{sample}



\section*{Acknowledgements}

This research has made use of the Astrophysics Data System, funded by NASA under Cooperative Agreement 80NSSC21M00561

\section*{Additional information}
The data were obtained from the publicly available NASA ADS database. All code used in this study is available at the project’s GitHub repository \href{https://github.com/vadibekyan/ads_stats}{ADS\_stats}.

\end{document}